\documentstyle[prl,aps,graphicx,amssymb]{revtex}
\begin{document}

\twocolumn[
\hsize\textwidth\columnwidth\hsize\csname@twocolumnfalse\endcsname

\title{Effect of the angular momentum on the magnitude of the current in magnetotunnelling spectroscopy of quantum dots}
\author{B.Jouault$^{1}$,J.P. Holder$^{2}$, M. Boero$^{2}$,G.Faini~\footnote{}$^{1}$,\\ \-F. Laruelle$^{1}$,E. Bedel$^{3}$, \- A.K. Savchenko$^{2}$,J.C. Inkson$^{2}$}
\address{
$^{1}$ L2M-CNRS, 196 Avenue H. Rav\'era, BP107, 92225 Bagneux Cedex, France \\
$^{2}$ Department of Physics, University of Exeter, Stocker Road, Exeter,EX4 4QL, UK \\
$^{3}$ LAAS-CNRS, 7 Avenue du colonel Roche, 31077 Toulouse Cedex, France \\
$*$e-mail:giancarlo.faini@L2M.CNRS.fr}

\maketitle
\begin{abstract}
We  present an experimental and theoretical study of electron tunnelling through quantum dots which focuses the attention on the  amplitude of the current resonances as a function of magnetic field. We demonstrate that the amplitudes of the resonances in the tunnelling spectra show a dramatically different  behaviour as a function of the  magnetic field, depending on the angular momentum of the dot state through which tunnelling occurs. This investigation allows us to probe the details of the confined wave functions of the quantum dot.
\end{abstract}


] 
Quantum dots (QDs) are structures where electrons are confined in all directions.  Ever since it has become possible to manufacture such structures there has been a considerable effort to develop spectroscopic tools capable of detecting the energy spectrum of QDs. Among the different techniques used to probe the structure of zero-dimensional systems, magneto-transport is one of the most common because the presence of a magnetic field introduces a structure-independent quantisation that is superimposed on the one produced by the fabrication process~\cite{bos92}.
In all these studies the emphasis has been placed on the shift of the I-V features produced by the QD states as a function of the magnetic field. This type of analysis has been performed to probe the QD spectrum, but it is insensitive to the details of the QD wave functions. One notable exception has been the study by P.H. Beton {\it et al.}~\cite{bet95} where the one dimensional wave functions were probed by means of the magneto-tunnelling.

In this letter we will perform an experimental and theoretical analysis of magneto-transport in vertical double-barrier systems enclosing a QD.  We shall focus our attention on the amplitude of the current resonances associated with the QD levels and on their variation with the applied magnetic field.
We have observed a striking difference in the behaviour of the amplitude of the current features. This is associated with the different angular momentum of the QD states through which the tunnelling occurs.

In order to interpret this effect,  we have developed a theoretical model which enables us to calculate the I-V characteristics of QD structures in the presence of a magnetic field analytically~\cite{jou98}. The model also shows that,
in the presence of the magnetic field, a selection rule applies which severely limits or suppresses altogether the current flowing through the QD states with an angular momentum co-linear to B.

\begin{figure}[!p]
\includegraphics*[width=3.09in]{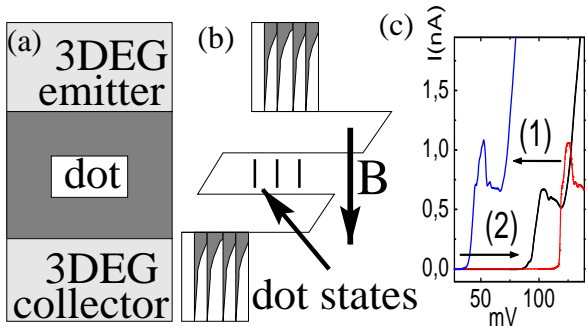}
\caption{\textbf{(a)} Schema of the device. \textbf{(b)} Schematic energy band diagram of the device under an applied bias. A magnetic field is applied along the current direction. \textbf{(c)} Shift of the threshold voltage with LED illumination (1) and warm-up (2). After each LED pulse, the whole spectrum permanently shifts towards lower voltage bias.}
\label{fig:schema}
\end{figure}

The structures considered in this work consist of two 3D contacts with a QD inserted in the middle of a $GaAlAs$-$GaAs$-$GaAlAs$ double barrier resonant tunnelling heterostructure ( see figures~\ref{fig:schema}a and~\ref{fig:schema}b). The fabrication of the samples is a combination of ion-implantation through metallic masks (used to define the quantum boxes lateral dimensions) patterned by electron beam and of epitaxial regrowth of the 3D $GaAs$ top electrode~\cite{fai96}. The $GaAs$ quantum well has a width of $5.1$nm; the well is embedded between two $Al_{0.33}Ga_{0.67}As$ barriers of width $8.7$nm. Si-doped contact layers with Si to $2\times 10^{17}cm^{-3}$ are formed on either side of the barriers. Care has been taken to prevent impurities inclusion in the quantum well region during the growth: the doped regions are separated from the barriers by a $20nm$ undoped spacer layer. These layers have been grown in two steps: the first $10nm$ at a temperature of $540~^{o}C$ to prevent Si segregation and the following $10nm$ plus the double barrier structure at $630~^{o}C$. 

\begin{figure}[!p]
\includegraphics[width=3in]{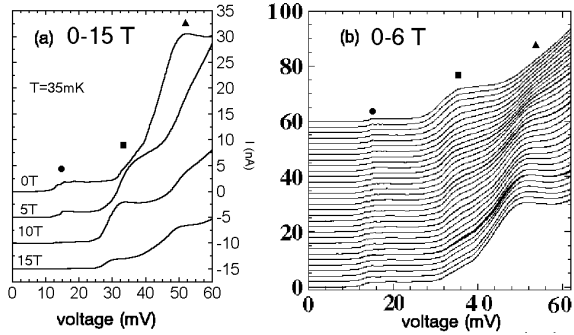}
\caption{\textbf{(a)} I-V characteristics of the 35nm-radius at $T=35 mK$ for different values of B. \textbf{(b)} Evolution of the I-V curve as B is increased from 0 to 6T in steps of 0.2T. All the curves have been offset for clarity.}
\label{fig:iv}
\end{figure}

In the following we will describe the results obtained on a nominal 35nm-radius dot at a temperature of $35mK$ and with a magnetic field applied along the vertical direction of the device, i.e. along the direction in which the current flows.  The main resonance current peak which is observed in larger devices is replaced by current plateaus ($\bullet$,$\blacksquare$,$\blacktriangle$), whose voltage positions and current amplitudes vary with B (see fig.~\ref{fig:iv})

Before we further discuss these plateaus, we briefly focus on the anomalously low threshold voltage (13mV), which is well below the expected value of 200mV.
In fig.~\ref{fig:schema}c we show the effect of LED illumination on the voltage position of the first current resonance at $T=35mK$. The I-V characteristics have been recorded after each LED pulse. We clearly observe a persistent shift of the whole spectrum to lower voltages, which can be removed by increasing the temperature~\cite{jou98b}.
We believe that this behaviour is related to the ion-implantation process used to define the QDs which induces the formation of deep charged defects under the lateral barriers. As in conventional 2DEG heterojunctions, the light changes the charge state of the deep defects, hence inducing a shift of the pinning of the Fermi level, lowering the threshold voltage. 
Despite the presence of such deep defects in the lateral barrier, we will show that the data presented here are unambiguously related to lateral confinement of the QD states.

\begin{figure}[!p]
\centering
\includegraphics*[width=3.5in]{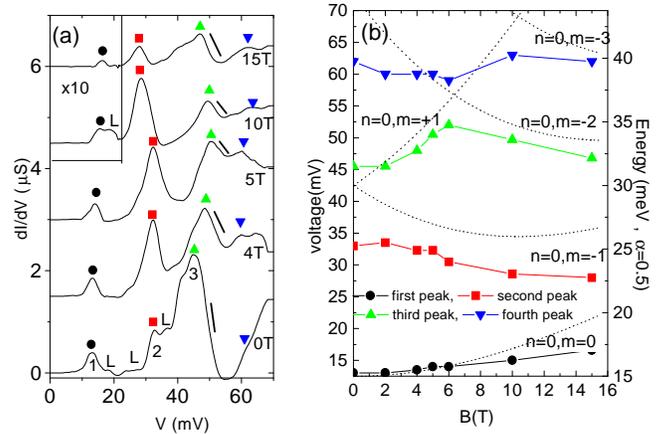}
\caption{\textbf{(a)} dI/dV spectra for several values of B from $0T$ to $15T$ \textbf{(b)} Voltage position of the dI/dV peaks as a function of B. The dotted lines represent the single-electron energy band diagram with $\alpha=0.5$ and \protect{$\hbar \omega_{0}=15meV$}.}
\label{fig:shifts}
\end{figure}

Let us first focus on the magnetic field dependence of the voltage positions of the current plateaus. In fig.~\ref{fig:shifts}a we show the $dI/dV$ curves corresponding to the I-V spectra: the differential conductance traces reveal another resonance at high bias ($\blacktriangledown$) in addition to the three main conductance peaks.
Minor peaks labeled with a $'L'$  are temperature independent and thus they are not related to QD states but to Local Density of State Fluctuations in the contacts~\cite{sch96}.
 
 The magnetic field displacement of the four conductance peaks with B is compared in fig.~\ref{fig:shifts}b with the single particle spectrum of the QD where the energy eigenvalues are given by the Darwin-Fock formula:
 \begin{equation}
\label{eqn:darwin}
 E_{n_{d},m_{d}}=\hbar\omega_{d}(2n_{d}+|m_{d}|+1)+\frac{1}{2}\hbar\omega_{c}m_{d}+E_{0}
 \end{equation} 
where $\omega_{d}=\sqrt{\omega_{0}^{2}+\frac{\omega_{c}^{2}}{4}}$, $\omega_{0}$ is the strength of the size confinement, $\omega_{c}$ is the cyclotron frequency and $E_{0}$ is the energy of the quantum well ground state in the absence of lateral confinement.

 The best fit for the position of the peaks on the voltage scale as a function of B gives $\hbar \omega_{0}=15meV$, with the voltage-to-energy conversion coefficient $\alpha= 0.5$. 
The radius of the first state is equal to $r \simeq \sqrt{2 \hbar/m^{*} \omega_{0}} \simeq 12 \mathrm{nm}$, which is in agreement with the nominal size of the QD taking into account the real lateral confinement potential profile induced by the ion-implantation inter-mixing process under the mask~\cite{vie92}. Although single donors in GaAs have a comparable radius ($10nm$), their excited states do not behave like the data of fig.~\ref{fig:shifts}b. Coulomb blockade is also irrelevant because the distance between the two first peaks decreases with B instead of slightly increasing since B confines more the electrons, hence increasing the Coulomb repulsion. Thus we are dealing with features releated to QD states for which a parabolic confinement and a single-electron picture are good approximations, given the strength of the size confinement in our structures~\cite{boe94}. Spin effects are also negleted here since we do not observe such effects in our experiments.

\begin{figure}[!p]
\centering  \includegraphics[width=3.5in]{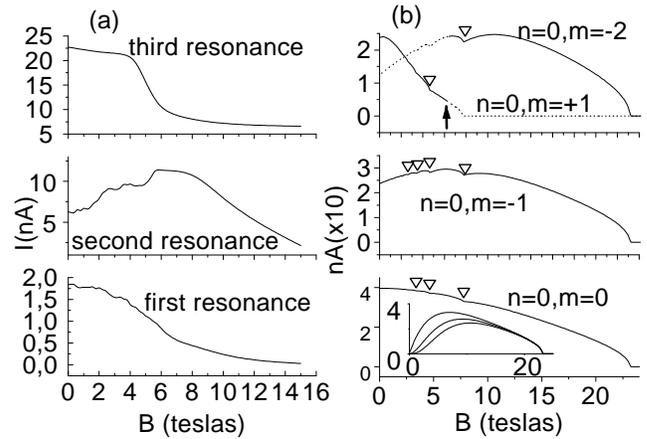}
\caption{\textbf{(a)}  Experimental current amplitude for the first three resonances of the 35nm-radius quantum dot device. The second and third resonances show, respectively, a maximum and a minimum at a magnetic field of about $6T$.
\textbf{(b)} Calculated current amplitude as a function of B for different quantum dot states. Parts of these curves which cannot correspond to the experimental third resonance are dotted.
\textbf{Inset}: Current contribution of the first Landau level in the emitter for $n_{e}=n_{d}=0$ and $m=0,-1,-2$ (from top to bottom).
}
\label{fig:ib}
\end{figure}

Let us focus now on the magnetic field dependence of the plateaus' current amplitudes. The fig.~\ref{fig:iv} shows that the amplitude of the first one decreases as B is increased, whereas that of the second one increases and then decreases. The third resonance amplitude decreases regularly. The current amplitudes of each QD state are calculated more precisely by integrating the conductance peaks in fig.~\ref{fig:shifts}a. The resulting curves have been reported on fig.~\ref{fig:ib}a for the first three main conductance peaks.
In order to explain their behaviour, we have developed a method to calculate the I-V characteristics as a function of B~\cite{jou98}. The QD eigenfunctions are:

\begin{equation}
 \Phi_{dot}(r,\varphi,z)=\frac{e^{im_{d}\varphi}}{\sqrt{2\pi}}
R_{n_{d},m_{d}}(r)\Psi_{d}(z)
 \end{equation}
where $R_{n_{d},m_{d}}(r)$ is a function that can be expressed in terms of the hypergeometric function \cite{lan,foc28} and $\Psi_{d}(z)$ is the ground state wave function of the one dimensional quantum well in the vertical direction. Similarly the applied magnetic field produces highly degenerate  1D Landau subbands in the contacts. The spectrum of the contacts is given by:

 \begin{equation}
 E_{n_{e},m_{e},k_{z}} =
 \hbar\omega_{e}(2n_{e}+m_{e}+|m_{e}|+1)+\frac{\hbar^{2}k^{2}_{z}}{2m^{*}}
 \end{equation}
 where $\omega_{e}=\omega_{c}/2$ and $k_{z}$ is the momentum in the growth direction while the eigenfunctions are plane waves along the vertical direction and for the lateral part they are of the same shape as those of the QD. The only difference is that in the contacts  only the magnetic field produces a quantisation whilst in the dot the effect of the magnetic field is superimposed on that of  the size confinement. Notice that the lowest Landau subband of energy $E=\hbar\omega_{e}$ does not present strictly positive values of $m_{e}$, the second subband of lateral energy $E=3\hbar\omega_{e}$ contains only the value  $m_{e} \leq 1$, and in general the $n_{e}$-th subband contains a maximum positive $m=n_{e}$. \\
Whenever the energy of a QD state falls between the Fermi level and
the conduction band bottom, tunnelling through such a state becomes energetically possible and current starts flowing through it. The amplitude of the current that flows is determined by the overlap between the QD wavefunctions and the electron wavefunctions in the contacts which are at the same energy.
The states in the contacts are highly degenerate 1D Landau subbands which are classified by the principal quantum number $n_{e}$ and the angular momentum $m_{e}$ along the direction of the field. The QD states are also classified in terms of a principal quantum number $n_{d}$ and the angular momentum $m_{d}$. It follows that the tunnelling process must obey a selection rule, i.e. tunnelling is only possible between states with the same angular momentum~\cite{jou98}. Therefore, of the many degenerate subbands contained in each Landau level in the contacts, only the one with the right value of angular momentum will contribute to the tunnelling through each QD state.

The overlap integrals~\cite{bar61} between the QD and the contacts' states can be evaluated analytically and for the lateral directions are given by:

 \begin{eqnarray}
 M_{n_{d},m_{d}:n_{e},m_{e}}^{\|}=\frac{\alpha_{d}\alpha_{e}}{2}\Gamma(\gamma)
 \lambda^{-n_{d}-n_{e}-\gamma}(\lambda-k_{d})^{n_{d}}
 \nonumber \\
 \times (\lambda-k_{e})^{n_{e}} F(-n_{d},-n_{e},\gamma,\frac{k_{d}k_{e}}{(\lambda-k_{d})(\lambda-k_{e})})\delta_{m_{e},m_{d}}
 \end{eqnarray}
 where $\alpha_{d,e}=\frac{1}{a_{d,e}^{1+|m|}}[\frac{(|m|+n_{d,e})!}
 {2^{|m|}n_{d,e}!|m|!^{2}}]^{\frac{1}{2}}$,
 $a_{d,e}=\sqrt{\frac{\hbar}{2m^{*}\omega_{e,d}(B)}}$, F is the
 hypergeometric function~\cite{lan},
 $\lambda=\frac{1}{4}(\frac{a^{2}_{d}+a^{2}_{e}}{a^{2}_{d} \times a^{2}_{e}})$,
 $k_{d,e}=\frac{1}{2a^{2}_{d,e}}$ and $\gamma=|m|+1$ (hereafter $m=m_{d}$ and $n=n_{d}$ for clarity).
 This formula can be combined with well known expressions for the overlap integral in the vertical direction~\cite{boe94,pay85} to obtain the total overlap between each dot state and the contacts and therefore calculate the current within a transfer matrix approach. This method is well known to correctly reproduce the physics even in the resonant tunnelling regime of double-barrier systems.

This technique has been adopted to calculate the I-V characteristics of 3D-0D-3D structures in the presence of an external magnetic field applied in the vertical direction, i.e. parallel to the current. Fig.~\ref{fig:ib}b shows the behaviour of the current amplitude for different states. States $n=0$ and $m=-2,-1,0,1$ are represented, calculated with the values of the parameters in accordance with those extracted from the fit using equation~\ref{eqn:darwin} of the experimental data. The amplitude for all states shows oscillations as a function of magnetic field. Some of them are marked by an open triangle. These oscillations have the characteristic $\frac{1}{B}$ dependence and are due to the oscillations in the density of states at the Fermi energy in the contacts as the magnetic field is changed.

As the magnetic field is increased, the resonances show a rather different behaviour in terms of their amplitude. The resonance associated with the ground state shows an amplitude that decreases systematically as a function of the magnetic field, while the first excited state $n=0,m=-1$ initially increases in amplitude, attains a maximum, and eventually decreases. The state $n=0,m=-2$ has the same qualitative features as the $n=0,m=-1$ one, but the maximum is more pronounced and shifted at higher magnetic field.

The presence of the magnetic field produces two effects. Firstly it reduces the momentum along the direction of the current for the electrons in the contacts by moving the subbands' minima to higher energies. Secondly it confines the electrons towards the centre of the dot. The first effect tends to diminish the current, while the second increases it by increasing the overlap in the lateral direction between the QD wavefunctions and those of the contacts. Thus each Landau subband in the emitter gives a current which attains a maximum at a finite magnetic field. The position of this maximum depends on the relative strength of the two effects. Moreover increasing the angular momentum $m$ decreases the overlap integral and shifts this maximum to higher field. In the inset of fig.~\ref{fig:ib}b we show the contribution of the first Landau level in the emitter for $m=0,-1,-2$ to clearly illustrate this shift of the maximum. The total current is obtained by summing the contribution of all the occupied
Landau levels in the emitter and it follows that the shift of the maximum current value from the origin B=0 T is a direct measurement of the magnitude of the angular momentum $m$.

Perhaps even more striking is the totally different behaviour of the states $n=0$,$m=1$ and $n=0$,$m=-1$ when B increases. At $B=0$ these states are degenerate and the radial part of the wave function is identical for both of them. When B is high enough ($\geq$ 7T), all electrons in the emitter are in the lowest Landau level ($n_{e}=0,m_{e} \leq 0$). These electrons can tunnel only into the QD states with the same non-positive angular momentum. As a result when $B \geq 7T$ tunnelling through the QD state $n_{d}=0,m_{d}=1$ is forbidden. As a general selection rule, tunnelling through a state with angular momentum $m$ is forbidden whenever there are only $m$ occupied Landau subbands in the contacts.

The figure~\ref{fig:ib} shows that our theory is in good agreement
with the experimental findings for the two first conductances peaks. However the complex theoretical behaviour of the third resonance is not observed. The theory predicts that the amplitude is given by the state $n=0, m=+1$ below $6T$ and by $n=0, m=-2$ at higher magnetic fields. Thus we expect to observe a decrease of the current as B increases up to $6T$, followed by a substantial increase and eventually a decrease at higher fields. Experimentally we observe only a decrease of the amplitude of this resonance. Nevertheless the coupling of the third and the fourth states ($\blacktriangle$ and $\blacktriangledown$) is clearly seen at $6T$ on the Darwin-Fock spectra (fig.~\ref{fig:shifts}b). Notice also that the right side slope of the third conductance peak (indicated by a bold line in fig.\ref{fig:shifts}a) is minimum at $6T$. So we attribute the experimental decrease of the third resonance amplitude to a poor overlap with the first Landau level in the emitter, which decreases the contribution to the current of the $n=0,m=-2$ state. Another explanation could be releated to the non parabolicity of the real QD.

\begin{sloppypar}
In conclusion, we have presented an analysis of the magneto-transport through QDs. We have shown that the current amplitude of the resonances is directly related to the angular momentum of the QD states contributing to the transport. The magnetic field strongly suppresses tunnelling through states with angular momentum parallel to it and in some cases tunnelling through such states is altogether forbidden. This analysis is substantiated by our experimental data collected on 3D-0D-3D structures and the agreement between the experimental findings and the theoretical results is very good.
The analysis can be readily  extended to 2D-0D-2D systems such as those where an accumulation layer is formed in the contacts~\cite{jou98} and, with more substantial modifications, to study QDs where the single particle picture no longer holds.

We wish to express our gratitude to C.~Mayeux and D.~Arquey for their technical help.
\end{sloppypar}


\end{document}